\documentclass[12pt,preprint]{aastex}

\shorttitle{Fast rotating subdwarf B}
\shortauthors{Geier et al.}

\begin{document}


\title{The fast rotating, low gravity subdwarf B star EC\,22081$-$1916 -- Remnant of a  common envelope merger event}


\author{S. Geier, L. Classen, U. Heber}
\affil{Dr. Karl Remeis-Observatory \& ECAP, Astronomical Institute, Friedrich-Alexander University Erlangen-Nuremberg, Sternwartstr.~7, D 96049 Bamberg, Germany}
\email{geier@sternwarte.uni-erlangen.de}



\begin{abstract}
Hot subdwarf B stars (sdBs) are evolved core helium-burning stars with very thin hydrogen envelopes. In order to form an sdB, the progenitor has to lose almost all of its hydrogen envelope right at the tip of the red giant branch. In binary systems, mass transfer to the companion provides the extraordinary mass loss required for their formation. However, apparently single sdBs exist as well and their formation is unclear since decades. The merger of helium white dwarfs leading to an ignition of core helium-burning or the merger of a helium core and a low mass star during the common envelope phase have been proposed. Here we report the discovery of EC\,22081$-$1916 as a fast rotating, single sdB star of low gravity. Its atmospheric parameters indicate, that the hydrogen envelope must be unusually thick, which is at variance with the He-WD merger scenario, but consistent with a common envelope merger of a low-mass, possibly substellar object with a red-giant core.
\end{abstract}





\section{Introduction}

Hot subdwarf B stars (sdBs) are evolved core helium-burning stars with very thin hydrogen envelopes residing at the extreme end of the blue horizontal branch \citep{heber09}. The formation of these stars remains unclear. Normal horizontal branch stars are formed after core helium-burning is ignited in the red-giant phase. Since the hydrogen envelopes of sdBs are extraordinarily thin, large mass-loss is necessary right at the tip of the red-giant branch. In the case of the close binaries among the hot subdwarfs -- about half of the known sdB stars are members of short-period (P $\lesssim$ 10 days) systems \citep{maxted01,napiwotzki04a} -- the required mass loss is triggered by the formation of a common envelope, which is finally ejected. The formation of sdBs with main sequence companions in wide orbits on the other hand can be explained by stable Roche lobe overflow \citep{han02,han03}. 

The formation of single sdBs is less well understood. Several single star scenarios are currently under discussion \citep[see][for a review]{heber09}, but all these scenarios require either a fine-tuning of parameters or extreme environmental conditions which are unlikely to be met for the bulk of the observed subdwarfs in the field. A particularly interesting scenario has been suggested by \citet{soker98} and \citet{nelemans98}. Accordingly, single sdB stars could be formed if a substellar companion in close orbit was engulfed by the red-giant progenitor and provided sufficient angular momentum to the common envelope before it was destroyed. 

Alternative scenarios invoke stellar mergers to form single sdB stars. The merger of binary white dwarfs was investigated by \citet{webbink84} as well as \citet{ibentutukov84} who showed that an EHB star can form when two helium core white dwarfs merge and the product is sufficiently massive to ignite helium. \citet{politano08} proposed the merger of a red giant and a low mass main-sequence star during the common envelope phase. However, the merger channel was under debate, because all single sdBs analysed so far turned out to be slow rotators \citep{geier09a} in contrast to the expectations. Here we report the discovery of the fast rotating, single sdB EC\,22018$-$1916.

\section{Observations}\label{obs}

EC\,22018$-$1916 ($V=12.9\,{\rm mag},\alpha_{2000}=22^{\rm h}10^{\rm m}52{\stackrel{\rm s}{\displaystyle .}}9$, $\delta_{2000}=-19^{\rm \circ}01'50''$) was discovered in the course of the Edinburgh-Cape blue object survey \citep{stobie97} and classified as an sdB star by \citet{copperwheat11}. Five high resolution spectra were taken with the FEROS spectrograph ($R=48\,000,\lambda=3750-9200\,{\rm \AA}$) mounted at the ESO/MPG\,2.2\,m telescope at La Silla. The first spectrum was taken on June 14, 2006 followed by two exposures on August 9 and 11 in the same year. The last two spectra were taken consecutively on October 30, 2010. In total, the data points cover a timespan of 4.5 years. The spectra have been reduced with the FEROS-DRS pipeline in the context of the MIDAS package. A median filter was applied to correct for cosmics. 

EC\,22018$-$1916 has been monitored by planetary transit surveys. Due to its favourable declination it was observed by both the All Sky Automated Survey \citep[ASAS,][]{pojmanski97} and the Northern Sky Variability Survey \citep[NSVS,][]{wozniak04}. Both light curves have been downloaded from the data archives. The ASAS dataset contains more than $450$ data points of Johnson-V photometry. We selected data extracted with smallest aperture and included only measurements of sufficient quality (flagged from A to C). The light curve was folded to a period of $0.15\,{\rm d}$ (see Sect.~\ref{binary}) and binned. The white light curve from NSVS contains $93$ data points and was phased in the same way. No variations exceeding $2\%$ (NSVS) and $1\%$ (ASAS) were detected. 

The Vizier database contains several consistent proper motion measurements of this object. Among them the PPMXL \citep[$\mu_{\rm \alpha}\cos{\delta}=17.0\,{\rm mas~yr^{-1}}$, $\mu_{\rm \delta}=-13.5\,{\rm mas~yr^{-1}}$,][]{roeser10}  and UCAC3 \citep[$\mu_{\rm \alpha}\cos\delta=17.7\,{\rm mas~yr^{-1}}$, $\mu_{\rm \delta}=-12.1\,{\rm mas~yr^{-1}}$,][]{zacharias10} values are independently measured and perfectly consistent within the error bars. Taking the average values we obtain $\mu=21.6\,{\rm mas~yr^{-1}}$. 

\section{Atmospheric parameters and rotational broadening \label{atmo}}

Atmospheric parameters and projected rotational velocity (see Table\,\ref{tab}) have been determined simultaneously by fitting a grid of synthetic spectra, calculated from line-blanketed, solar-metalicity LTE model atmospheres \citep{heber00}, to the hydrogen Balmer lines (H$_{\rm \beta}$-H$_{\rm 10}$) and helium lines (He\,{\sc i}\,4026,4472,4922,5876\,${\rm \AA}$ and He\,{\sc ii} 4686\,${\rm \AA}$) using the SPAS routine developed by H. Hirsch \citep[e.g,][]{geier11b}. The single spectra have been corrected for their orbital motion and coadded. Statistical errors are determined with a bootstrapping algorithm. 

\begin{figure}[t!]
\begin{center}
	\resizebox{16cm}{!}{\includegraphics{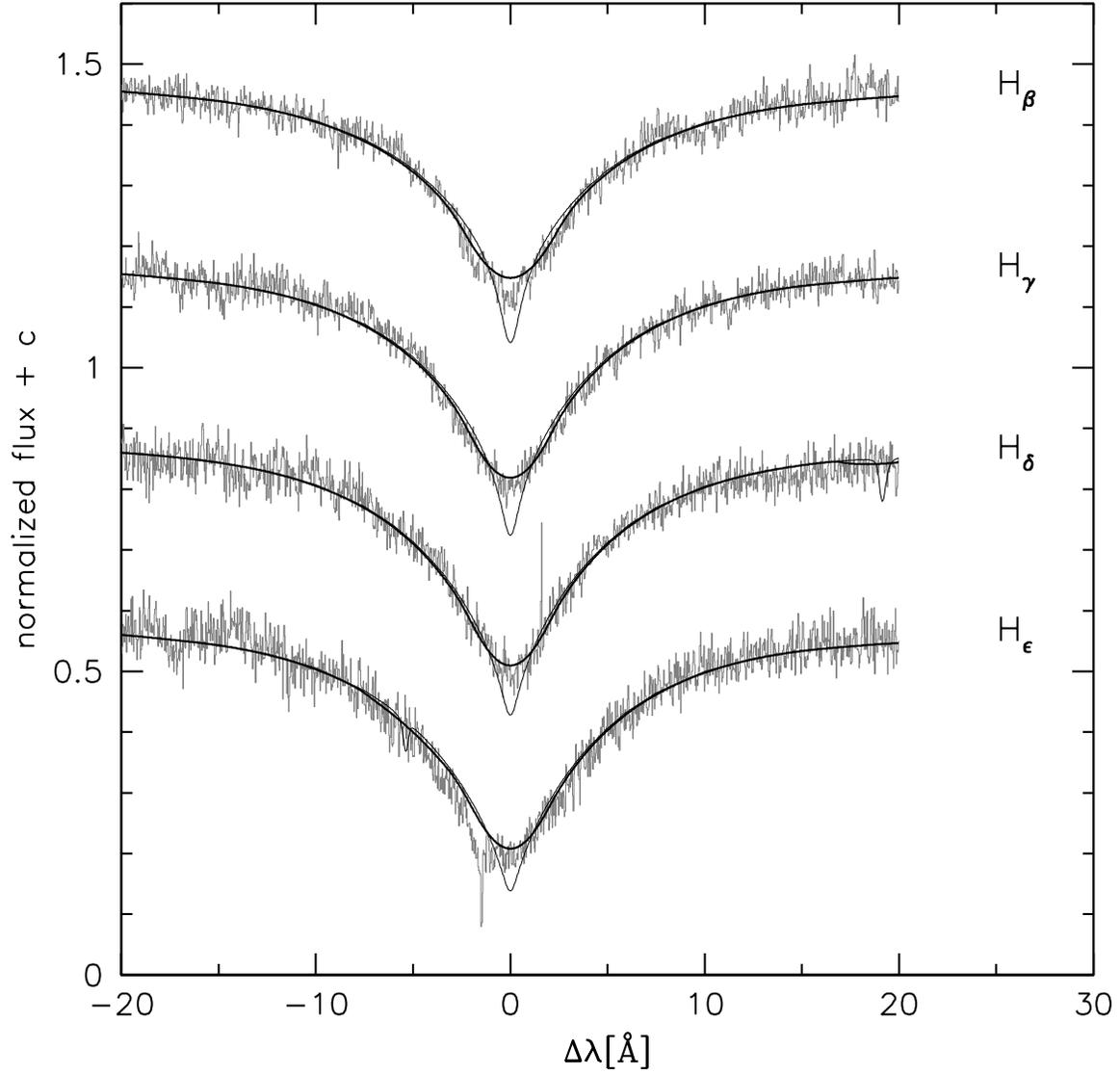}}
\end{center}
\caption{Fit of synthetic LTE models to some hydrogen Balmer lines. The thin solid line marks models without rotational broadening, the thick solid line the best fitting model spectrum with $v_{\rm rot}\sin{i}=163\,{\rm km\,s^{-1}}$.}
\label{fitH}
\end{figure}

\begin{figure}[t!]
\begin{center}
	\resizebox{16cm}{!}{\includegraphics{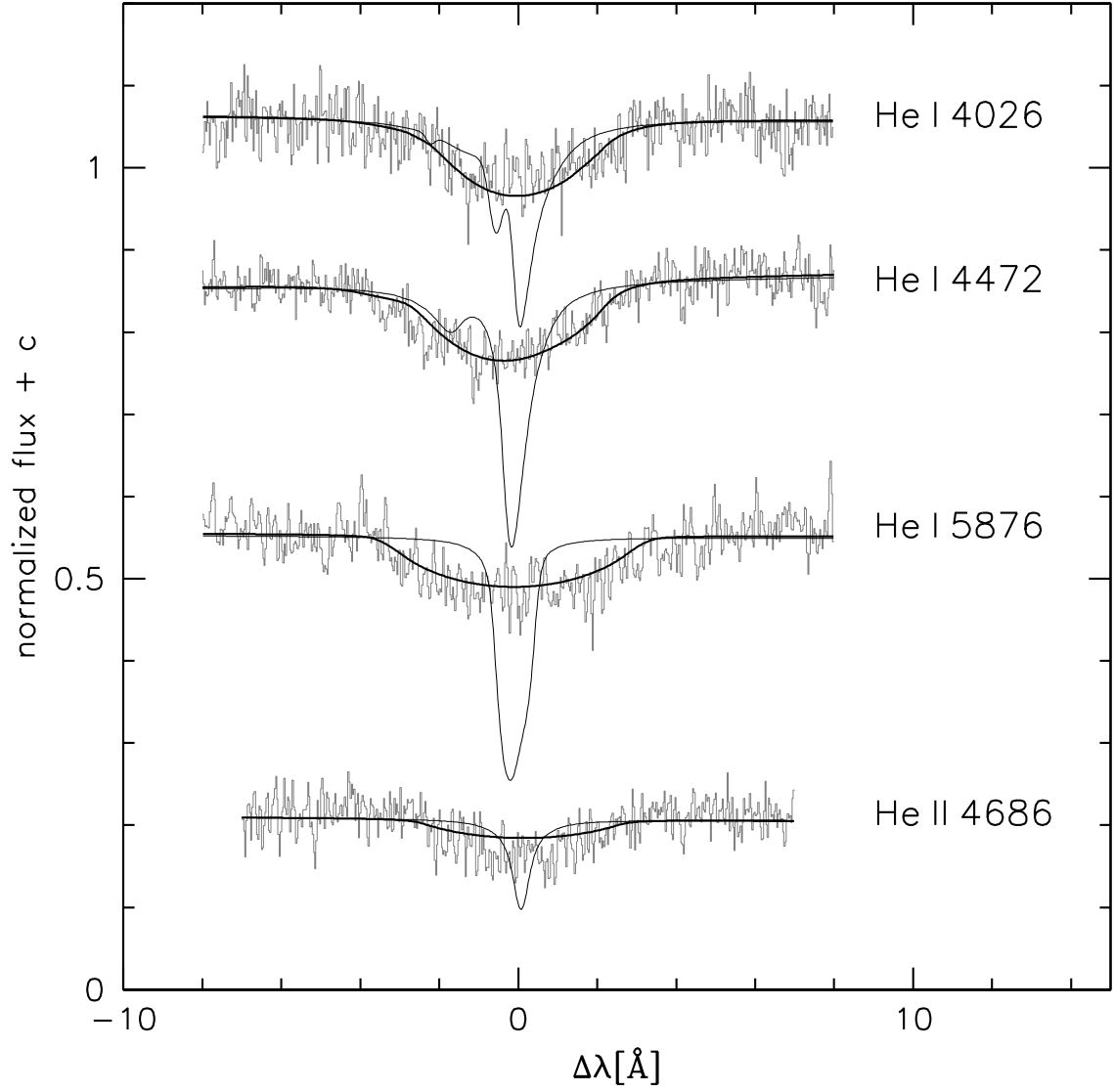}}
\end{center}
\caption{Fit of synthetic LTE models to helium lines (see Figure~\ref{fitH}). The extreme rotational broadening of the lines is obvious.}
\label{fitHe}
\end{figure}

As can be seen in Figures~\ref{fitH} and \ref{fitHe}, the Balmer line cores and the helium lines are significantly broadened. In order to fit these lines, a very high rotational broadening of $v_{\rm rot}\sin{i}=163\pm3\,{\rm km\,s^{-1}}$ is necessary. The fact that no metal lines have been found is consistent with this high broadening since weak features melt into the continuum in this case. The resulting effective temperature $T_{\rm eff}=31\,100\pm1000\,{\rm K}$,  and helium abundance $\log{y}=-1.97\pm0.02$ are typical for sdB stars, whereas the surface gravity of $\log{g}=4.77\pm0.10$ is unusually low for the effective temperature in question (see Figure~\ref{tefflogg}).

The radial velocities of the five single spectra were measured by fitting model spectra with fixed parameters derived from the spectral analysis to the Balmer and helium lines using the FITSB2 routine \citep[][see Table\,\ref{tab}]{napiwotzki04b}. No significant RV variations were measured. The radial velocity of the star is constant at $-13.1\pm3.6\,{\rm km\,s^{-1}}$. 

By adopting the canonical mass of sdB stars ($0.47\,M_{\rm \odot}$) we can derive the distance from the atmospheric parameters and the apparent magnitude following \citet{ramspeck01}. The transversal velocity $v_{\rm t}$ of the star in ${\rm km\,s^{-1}}$ can then be calculated using the simple formula $v_{\rm t}=4.74d\mu$, where the distance $d$ is given in kpc and the proper motion $\mu$ in mas. The distance to the star is $\simeq1.5\,{\rm kpc}$ and the transversal velocity $\simeq150\,{\rm km\,s^{-1}}$ perfectly consistent with an evolved star in the thick disk or in the halo \citep[e.g.][]{tillich11}. 

\section{Constraining the nature of EC\,22018$-$1916}

EC\,22018$-$1916 has the highest $v_{\rm rot}\sin{i}$ ever measured for an sdB star. All other single sdB stars analysed so far have $v_{\rm rot}\sin{i}<10\,{\rm km\,s^{-1}}$ \citep{geier09a}. In the following we discuss and exclude several possible explanations for this finding.

\paragraph*{Main sequence star?} 

Rotational velocities exceeding $100\,{\rm km\,s^{-1}}$ are quite common among main sequence A and B stars. Since the surface gravity $\log{g}=4.77$ is at the lower end of the hot subdwarf parameter range, the star may be regarded as a missclassified massive main sequence star. This interpretation, however, can be ruled out because the surface gravity is too high (see Figure~\ref{tefflogg}) and the helium abundance ($1/10-$solar) far to low. A double-lined binary consisting of two hot main sequence stars is another option. An unresolved, double-lined binary may explain the high measured surface gravity, which could be overestimated in this case. However, we cannot imagine a combination of main sequence stars, which would produce such an unusual spectrum.

\paragraph*{Hot subdwarf with unresolved pulsations?}

Two kinds of sdB pulsators are known. The slow pulsations of the V\,1093\,Her stars (sdBV$_{\rm s}$) are not expected to influence the line broadening significantly. In the case of the short-period pulsators (V\,361\,Hya type, sdBV$_{\rm r}$), unresolved pulsations can severely affect the broadening of the lines and therefore mimic higher ${v_{\rm rot}\sin\,i}$. \citet{telting08} showed that this happens in the case of the hybrid pulsator Balloon\,090100001. Unresolved pulsations are also most likely responsible for the high ${v_{\rm rot}\sin\,i}=39\,{\rm km\,s^{-1}}$ measured for the high-amplitude pulsator PG\,1605+072 \citep{heber99}.

\begin{figure}[t!]
\begin{center}
	\resizebox{16cm}{!}{\includegraphics{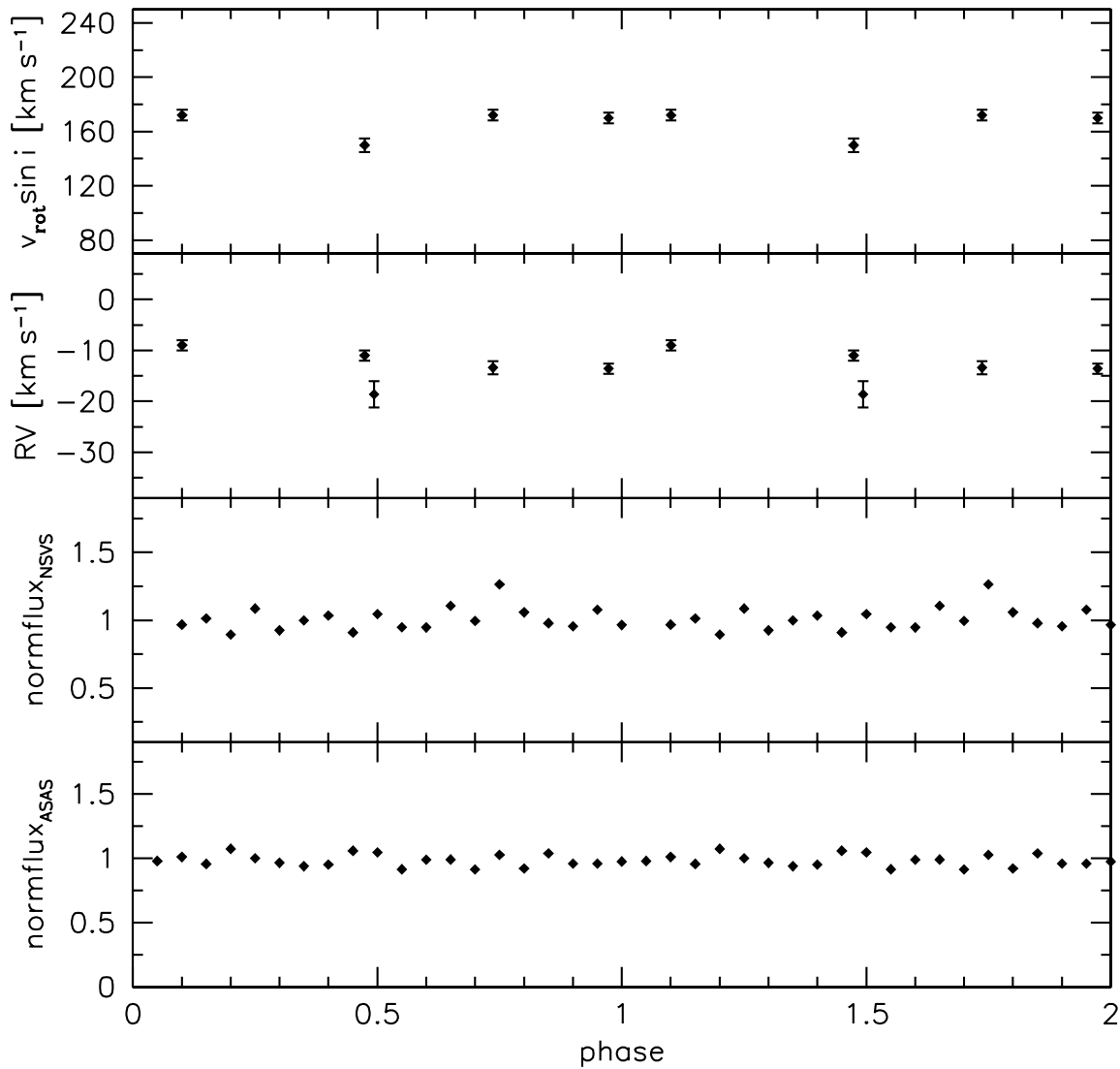}}
\end{center}
\caption{Time resolved photometry (ASAS V-band, NSVS white light, lower panels), radial and projected rotational velocities (with formal $1\sigma$ fitting errors, upper panels) plotted against phase. The data has been folded to the upper limit for the orbital period of a putative close binary with substellar companion ($P\simeq0.15\,{\rm d}$, see Sect.~4). Two complete phases are plotted for better visualisation. Since no significant variations can be seen in the data, EC\,22018$-$1916 is neither a strong pulsator nor in a close binary system.}
\label{nondetect}
\end{figure}

The five spectra of EC\,22018$-$1916 have been taken with exposure times ranging from $900\,{\rm s}$ to $1500\,{\rm s}$. The effective temperature of the sdB is consistent with the ones of short-period pulsating sdBs. The typical pulsation periods of sdBV$_{\rm r}$ stars are of the order of a few minutes and therefore shorter than the exposure times. However, the measured ${v_{\rm rot}\sin\,i}=163\,{\rm km\,s^{-1}}$ is so high that very large photometric variations at periods of a few minutes would be inevitable. Therefore the broadening cannot be caused by unresolved pulsations. The prominent mode of the strongest known sdB pulsator PG\,1605+072 has a photometric amplitude of $\simeq13\%$ \citep{koen98} and an RV amplitude of $\simeq15\,{\rm km\,s^{-1}}$ \citep{otoole05a}. In order to cause the line broadening necessary to fit EC\,22018$-$1916, both values would have to be much higher. Due to the fact that we neither detect RV variations nor any features in the light curves (see Figure~\ref{nondetect}) we conclude that EC\,22018$-$1916 is not a high-amplitude pulsator.

\paragraph*{Hot subdwarf with high magnetic fields?} 

\citet{otoole05b} discovered magnetic fields up to $\simeq1.5\,{\rm kG}$ in a small sample of sdB stars. From the analysis of magnetic white dwarfs with field strengths in the MG range it is known that small Zeeman splitting can mimic a broadening of the spectral lines \citep[see e.g.,][]{kuelebi09}. Could EC\,22018$-$1916 be the prototype of a new sdB class with very strong magnetic fields? Looking at Figs.~\ref{fitH} and \ref{fitHe} this explanation can be ruled out as well. Zeeman splitting affects every single spectral line in a different way. In contrast to that, the broadening of the lines is uniform. We therefore conclude that the line broadening of EC\,22018$-$1916 is caused by rotation.

\paragraph*{Close binary with large RV amplitude and orbital smearing?}

The hot subdwarfs with the highest measured projected rotational velocities ($v_{\rm rot}\sin{i}>100\,{\rm km\,s^{-1}}$) all reside in very close binary systems with orbital periods of $\simeq0.1\,{\rm d}$. These sdBs were spun up by the tidal influence of their close companions and their rotation became synchronised to their orbital motion \citep{geier07, geier10}. 

A close companion would therefore be the most natural explanation for the high $v_{\rm rot}\sin{i}$ of EC\,22018$-$1916. The colours of this star \citep[$J-K_{S}\simeq0.0$, 2MASS; ][]{skrutskie06} do not show any signs of a cool companion \citep{stark03}. A possible unseen companion must therefore be either a low mass main sequence star, a compact object like a white dwarf or a substellar object. 

White dwarf and main sequence companions can be immediately ruled out because no significant RV variations are detected on timescales of years, days and half an hour. This is also a strong argument against the hypothesis, that the strong line broadening may be at least partly caused by orbital smearing. Since the exposure times of the FEROS spectra are rather long ($900-1500\,{\rm s}$), the RV shift during the exposure would be large in a close binary with high RV amplitude \citep[e.g.][]{geier07}. However, no RV variations are measured (see Figure~\ref{nondetect}, upper panel). Furthermore, a quantitative spectral analysis as outlined in Sect.~\ref{atmo} has been performed for the single FEROS spectra and no significant variations in the atmospheric parameters or the $v_{\rm rot}\sin{i}$ were detected (see Table~\ref{tab}). Orbital smearing can therefore be ruled out. 

\paragraph*{Close binary with small RV amplitude?}\label{binary}

The remaining option would be a substellar companion in a very close orbit similar to the sdB+BD binary SDSS\,J08205+0008 \citep{geier11a}. Assuming that the rotation of EC\,22018$-$1916 is synchronised an upper limit for the orbital period can be calculated. Adopting $M_{\rm sdB}=0.47\,M_{\rm \odot}$ and using the measured $\log{g}$ the radius of the star is $R_{\rm sdB}=\sqrt{M_{\rm sdB}G/g}\simeq0.47\,R_{\rm \odot}$. Taking the inclination into account ($v_{\rm rot}\ge v_{\rm rot}\sin{i}$) we can calculate an upper limit for the orbital period $P\le2\pi R_{\rm sdB}/v_{\rm rot}\sin{i}\simeq0.145\,{\rm d}$. Another strict constraint is set by the lack of significant RV variations. Taking the standard deviation of the RV measurements and multiplying it by three we end up with a conservative upper limit for the RV semiamplitude $K<12\,{\rm km\,s^{-1}}$. 

Adopting the upper limits for $P$ and $K$, the companion would have to be a brown dwarf with $\simeq20\,M_{\rm J}$ and a radius of $\simeq0.1\,R_{\rm \odot}$. For inclinations lower than $90^{\rm \circ}$ the orbital period of the putative binary must be shorter, because  the absolute rotational velocity of the sdB has to be higher to keep $v_{\rm rot}\sin{i}$ fixed at the observed value. 

However, other important constraints have to be met as well. Neither the sdB nor its putative companion are allowed to fill their Roche lobes, because in this case the system would exchange mass. Since no indicative features for ongoing mass transfer (e.g. emission lines) are present in the spectra, the system must be detached. Calculating the Roche radii of both components as outlined in \citet{eggleton83} we derive a minimum orbital period for the system of $\simeq0.11\,{\rm d}$ and a minimum inclination of $47^{\rm \circ}$. For shorter periods and hence lower inclinations the sdB would fill its Roche lobe. A similar limit ($\simeq0.1\,{\rm d}$) is derived, if we allow $K$ to be smaller than $12\,{\rm km\,s^{-1}}$. For orbital periods shorter than that, a putative brown dwarf companion would fill its Roche lobe.

These simple calculations show that the possible parameter space of a close and synchronised binary would be extremely narrow ($P\simeq0.1-0.15\,{\rm d}$, $K\simeq4-12\,{\rm km\,s^{-1}}$). Furthermore, all possible configurations would lead to photometric variabilities easily visible in the light curve. Close sdB+dM or BD systems are not only often eclipsing, but also show sinusoidal variations due to light from the irradiated surface of the cool companion \citep[e.g.][]{oestensen10,for10,geier11a}. Due to its high temperature and low surface gravity EC\,22018$-$1916 has a very high luminosity compared to other sdBs, which should lead to a very strong reflection effect at inclinations of $\simeq50^{\rm \circ}$ or higher. Since no variations were found in the ASAS and NSVS light curves (Figure~\ref{nondetect}), a nearby low-mass companion can be excluded as well.

\section{Conclusion}

\begin{figure}[t!]
\begin{center}
	\resizebox{16cm}{!}{\includegraphics{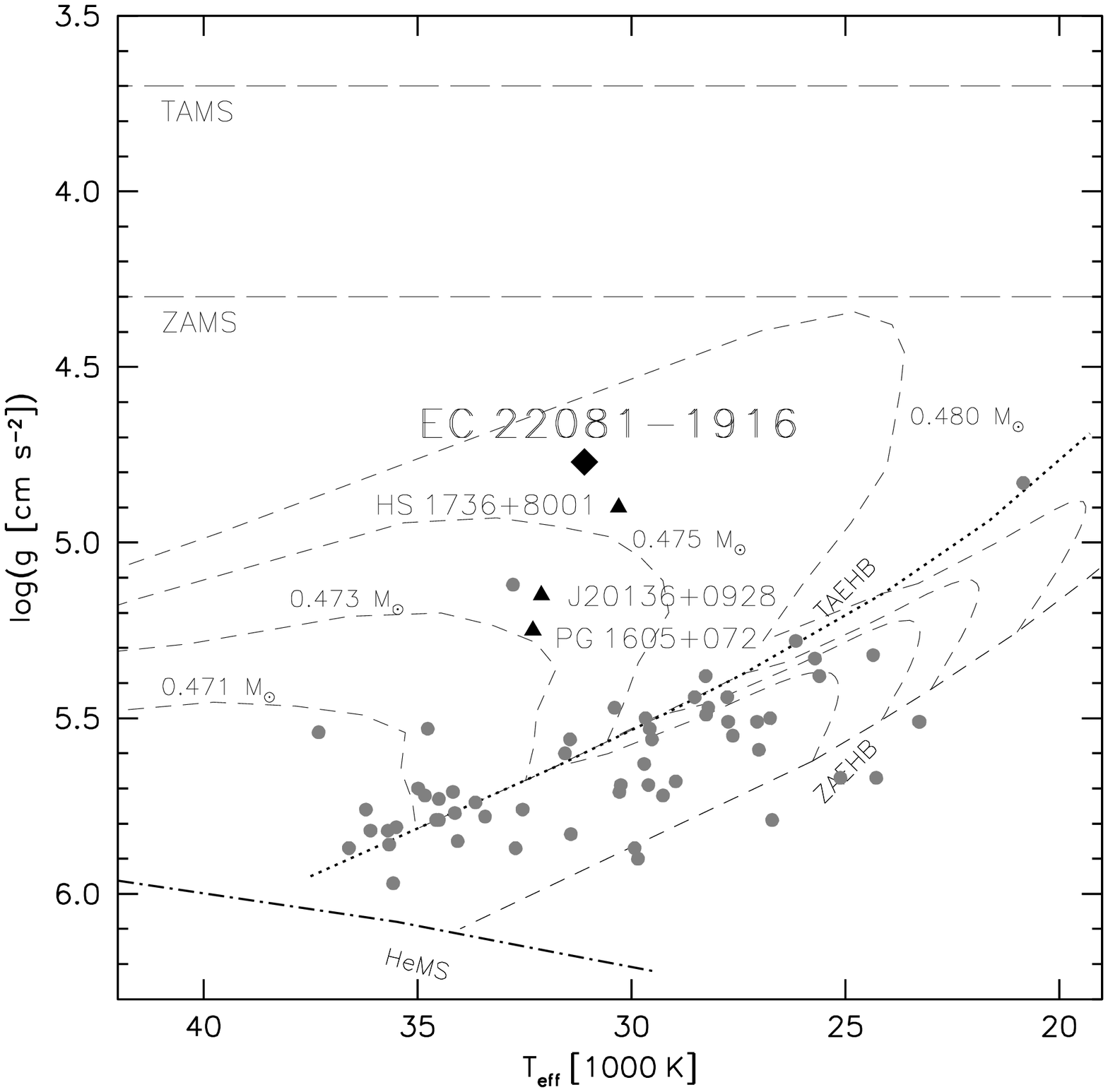}}
\end{center}
\caption{$T_{\rm eff}-\log{g}$ diagram. The grey circles mark sdBs from the SPY project \citep{lisker05}. The low gravity sdBs HS\,1736+8001 \citep{edelmann03}, PG\,1605+072 \citep{heber99} and J20136+0928 \citep{oestensen11} are plotted as triangles. The helium main sequence (HeMS) and the EHB band (limited by the zero-age EHB, ZAEHB, and the terminal-age EHB, TAEHB) are superimposed with EHB evolutionary tracks for solar metallicity from \citet{dorman93}. The location of the main sequence (MS, limited by the zero-age MS, ZAMS, an the terminal-age MS, TAMS) is indicated by the long-dashed horizontal lines.}
\label{tefflogg}
\end{figure}

After excluding all possible alternative scenarios we conclude that EC\,22018$-$1916 is the first single sdB star which is rapidly rotating.
Furthermore, the $\log{g}$ of EC\,22018$-$1916 is the lowest one ever measured for an sdB (see Figure~\ref{tefflogg}). \citet{oestensen11} argued that the low gravity of the pulsating sdB J20163+0928 ($\log{g}=5.15$) may be due to a rather thick layer of hydrogen. In the model of \citet{han02,han03} even the merger remnants with the highest masses would need a hydrogen layer of $\simeq0.01\,M_{\rm \odot}$ to reach at such low surface gravities.

The formation of such an object through single star evolution is very hard to explain. EC\,22018$-$1916 thus might have been formed by a merger event. Three merger scenarios have been proposed to explain the origin of hot subdwarfs. \citet{webbink84} and \citet{ibentutukov84} proposed the merger of two He-WDs as possible formation channel, which has been further explored by \citet{saio02}. \citet{han02,han03} included this channel in their binary evolution calculations and were able to model both the UV-excess in elliptical galaxies \citep{han07} and the different close binary fractions of sdBs in populations of different age in a consistent way \citep{han08}. He-WD mergers are believed to have very small envelope masses and are expected to be situated at the very blue end of the EHB. Both is at variance with the position of EC\,22018$-$1916 in the ($T_{\rm eff}-\log{g}$)-diagram (see Figure~\ref{tefflogg}). \citet{justham10} proposed that the merger of a close binary system consisting of an sdB and a He-WD may form a single helium enriched sdO. EC\,22018$-$1916, however, is helium-deficient.

EC\,22018$-$1916 most likely belongs to an old stellar population, either thick disk or halo. Its position in the ($T_{\rm eff}$-$\log{g}$)-diagram (see Figure~\ref{tefflogg}) may indicate a mass higher than canonical, which would be consistent with the predictions by \citet{han02,han03}. If it should be the remnant of a He-WD merger, this would imply important constraints on the merger process itself. Since the helium abundance of EC\,22018$-$1916 is ten times below the solar value, enough hydrogen must have survived the merger and must have been enriched in the atmosphere by diffusion processes. 

The third channel was suggested by \citet{soker98} and further explored by \citet{soker00,soker07}. \citet{politano08} followed this idea and focused on the formation of hot subdwarfs. The merger of a red-giant core and a low-mass, main-sequence star or substellar object during a common envelope phase may lead to the formation of a rapidly rotating hot subdwarf star. This scenario fits particularly well with observations for several reasons. 

First, the helium core of a red giant merges with an unevolved low-mass star or a brown dwarf. Both have hydrogen-rich envelopes. This provides a natural explanation for the low He abundance and surface gravity of the remnant. The hydrogen is provided by the merged companion. Furthermore, this companion also provides the energy required to eject the envelope and form the sdB. Several sdBs with low mass stellar and substellar companions have been found most recently and the true number may be much higher due to selection effects \citep[e.g.][]{for10,oestensen10,geier09b,geier11a,geier11c}. 

A very important predicition made by \citet{politano08} is that sdBs formed via the CE-merger channel should be rare. EC\,22018$-$1916 is unique among $\simeq100$ slowly rotating sdB stars analysed so far \citep{geier09a}. In contrast to that, \citet{han02,han03} predict a large fraction if not all of the single sdBs to be formed by WD-mergers. Unless there is a mechanism to get rid off all the angular momentum involved in a merger as suggested by \citet{saio02}, this observation is hard to explain. 

Furthermore, \citet{politano08} predict that a large fraction of the sdBs formed after CE-merger should rotate with a critical velocity $v_{\rm crit}$, which is defined as the rotational velocity at which mass loss induced by centrifugal forces prevents the red-giant core to accrete more material from the secondary. \citet{politano08} estimate this critical velocity to be about one third of the breakup velocity $v_{\rm br}=(GM/R)^{1/2}$. Using the parameters derived for EC\,22018$-$1916, we calculate $v_{\rm crit}\simeq145\,{\rm km\,s^{-1}}$ perfectly consistent with the projected rotational velocity measured from the spectrum.

In conclusion, the scenario proposed by \citet{soker98} and \citet{politano08} fits best with the observational data obtained so far, although the He-WD+He-WD or sdB+He-WD merger scenarios cannot be ruled out. EC\,22018$-$1916 is the first candidate for a merger remnant among the hot subdwarf stars. Similar objects are expected to be found in large spectroscopic databases like SDSS. Due to the high rotational broadening the quality of these data should be sufficient to find them.




\acknowledgments

Based on observations at the La Silla Observatory of the 
European Southern Observatory for programmes number 082.D-0649 and 084.D-0348. S.~G. is supported by the Deutsche Forschungsgemeinschaft (DFG) through grant HE1356/49-1. We thank L. Morales-Rueda for sharing her data with us. Furthermore, S.~G. wants to thank Ph. Podsiadlowski, C.~S. Jeffery, R.~H. \O stensen and S.~J. O'Toole for defending the merger channel as possible formation scenario for hot subdwarfs. Special thanks go to the organizers of the 4th sdOB meeting in Shanghai where these and other problems were discussed.

\clearpage



\begin{table}
\caption{\bf Parameters of EC\,22081$-$1916}
\label{tab}
\begin{center}
\begin{tabular}{lrrrrr}
\tableline
\noalign{\smallskip}
mid-HJD & RV & $T_{\rm eff}$ & $\log{g}$ & $\log{y}$ & $v_{\rm rot}\sin{i}$ \\
 & [${\rm km~s^{-1}}$] & [K] & & & [${\rm km~s^{-1}}$] \\
\noalign{\smallskip}
\tableline
\noalign{\smallskip}
2453900.845477 &  $-11.0\pm1.0$ & $31\,500\pm150$ & $4.74\pm0.02$ & $-1.91\pm0.03$ & $172\pm4.0$ \\
2453956.798348\tablenotemark{a} & $-18.6\pm2.6$ & $-$ & $-$ & $-$ \\
2453958.634893 &  $-13.4\pm1.3$ & $31\,500\pm190$ & $4.81\pm0.03$ & $-1.83\pm0.04$ & $150\pm5.0$ \\
2455499.620294 &  $-13.6\pm1.0$ & $29\,800\pm140$ & $4.67\pm0.02$ & $-2.01\pm0.03$ & $172\pm4.0$ \\
2455499.639508 &  $-9.0\pm1.0$ & $31\,200\pm170$ & $4.83\pm0.03$ & $-1.90\pm0.04$ & $170\pm4.0$ \\
\noalign{\smallskip}
\tableline
\noalign{\smallskip}
Coadded & & $31\,100\pm100$ & $4.77\pm0.02$ & $-1.97\pm0.02$ & $163\pm3.0$ \\ 
\noalign{\smallskip}
\tableline
\end{tabular}
\end{center}
\tablenotetext{a}{Spectrum with low S/N.}
\end{table}

\clearpage

\clearpage




\end{document}